# In-Situ Polarimetry in Collimated Magneto-Infrared Spectroscopy System

Zeping Shi[1†], Wenbin Wu[1†], Zhiwei Zhang[1†], Yuhan Du[1†], Chenyao Xu[1†], Congming Hao[1], Xiangyu Jiang[1], Xin Chen[1], Guangyi Wang[1], Mingsen Zhou[1], Chunhui Pan[2], Wei Lu[2], Hao Shen[1], Haifeng Pan[1], Zhenrong Sun[1], Junhao Chu[3,4], Xiang Yuan[1,3,5*]

[1] State Key Laboratory of Precision Spectroscopy, East China Normal University, Shanghai 200241, China

[2] Multifunctional Platform for Innovation Precision Machining Center, East China Normal University, 200241, Shanghai, China

[3] School of Physics and Electronic Science, Key Laboratory of Polar Materials and Devices, Ministry of Education, East China Normal University, Shanghai 200241, China

[4] Institute of Optoelectronics, Fudan University, 200438, Shanghai, China

[5] Wuhan National High Magnetic Field Center, Huazhong University of Science & Technology, Wuhan, China, 430074

[†]These authors contributed equally to this work.

[*]Correspondence and requests for materials should be addressed to X. Y. (E-mail: xyuan@lps.ecnu.edu.cn)

**Abstract**

Magneto-infrared spectroscopy under strong magnetic fields provides a powerful probe of Landau quantization and field-induced collective excitations, yet its full potential has long been constrained by the lack of in-situ polarization control, because the highly divergent infrared beam propagating through narrow light tubes undergoes multiple wall reflections, leading to severe polarization degradation. Here we report a collimated magneto-infrared spectroscopy system that integrates continuous in-situ polarimetry. The system employs incident and exit collimation chambers forming a Kepler type optical architecture, which converts the large-aperture FTIR output into a low-divergence beam and strongly suppresses multi-reflection trajectories inside long gold-plated light tubes, thereby enhancing both optical throughput and polarization fidelity. A remotely controlled polarization module, consisting of an automated linear polarizer and a switchable Fresnel rhomb positioned entirely outside the high-field region, enables continuous in-situ tuning between linear, circular, and arbitrary elliptical polarization states without thermal cycling, manual realignment, or breaking vacuum. Interchangeable compact focusing modules further support Faraday and Voigt geometries in both transmission and reflection experiments within a 50 mm magnet bore, providing efficient beam focusing and signal collection while maintaining polarization fidelity. The setup achieves a minimum root-mean-square noise of 0.0033%, an average noise of 0.0082%, and a linear polarization extinction ratio up to 40:1. We demonstrate the capability through continuous in-situ linear polarimetry and broadband circular polarimetry in the magneto-infrared spectroscopy of various single crystals. This platform establishes a robust experimental framework for in-situ polarization-resolved magneto-infrared spectroscopy.

## I. INTRODUCTION

Infrared spectroscopy has long been a cornerstone technique across physics, astronomy, chemistry, and materials science, owing to its unique capability to probe elementary excitations and low-energy electronic structures[1–8]. In condensed matter physics in particular, infrared spectroscopy provides direct access to lattice vibrations, carrier dynamics, collective modes, and electronic band structures through frequency-resolved optical responses such as absorption and reflectivity. By analyzing these frequency-dependent optical properties, one can quantitatively extract dielectric functions and optical conductivities, thereby revealing microscopic information including band dispersion, energy gaps, and interaction-induced renormalizations[9–14].

Infrared spectroscopy primarily exploits information encoded in the photon frequency dimension. A complete optical characterization of matter generally requires access to the polarization degree of freedom. Polarization-resolved infrared spectroscopy enables selective coupling to specific electric-dipole matrix elements, allowing different components of the dielectric function tensor to be disentangled. This capability is essential for identifying symmetry, quantifying anisotropic charge transport, resolving magnetoelectric coupling, and tracking structural or electronic phase transitions[15–21]. In the absence of polarization resolution, many symmetry-sensitive optical responses are effectively averaged out, obscuring key physical information.

When polarization control is combined with high magnetic fields, magneto-infrared spectroscopy becomes an exceptionally powerful probe of quantum matter[22–27]. Under sufficiently strong magnetic fields, charge carriers in clean insulating or semimetallic systems undergo Landau quantization, giving rise to discrete inter-Landau-level transitions that are highly sensitive to band topology and carrier statistics[28]. These transitions obey strict polarization-dependent optical selection rules. In the Faraday geometry, where the magnetic field is parallel to the light propagation direction, inter-Landau-level transitions are typically locked to circular polarization[29]. In contrast, in the Voigt geometry, where the magnetic field is perpendicular to the light propagation direction, the selection rules are predominantly governed by linear polarization[30,31]. Polarization-resolved magneto-infrared spectroscopy therefore allows direct experimental access to Landau-level wavefunction properties[29], such as parity, which are difficult or impossible to obtain using transport or tunneling probes alone. Beyond Landau quantization, polarization-resolved magneto-infrared measurements enable direct investigations of a wide range of magnetic-field-induced quasiparticle excitations. For instance, the polarization response of infrared-active phonons can be used to determine the orientation and evolution of phonon charge under magnetic fields, providing insight into dynamical magnetoelectric effects[32]. Circularly polarized magneto-infrared measurements further allow the extraction of Faraday and Kerr rotations[33], offering direct detections of magnetic order parameters and, in certain topological systems, optical identification of topological invariants. Over the past decades, large-scale facilities such as the National High Magnetic Field Laboratory and the European Magnetic Field Laboratory, as well as laboratory-scale superconducting magnet platforms[34–44], have generated polarization-sensitive magneto-infrared studies, significantly advancing research in topological and correlated quantum materials.

Despite these advances, existing polarization-resolved magneto-infrared Fourier-transform spectroscopy systems generally lack in-situ polarimetry capability. This limitation leads to several fundamental experimental challenges. First, polarization switching typically requires reconfiguring optical components or resetting the optical path, making it difficult to quantitatively compare magneto-infrared responses of the same sample under different polarization states. Second, most systems are restricted to a few discrete polarization configurations, preventing continuous polarization modulation or scanning. Third, the absence of reliable in-situ polarization control hinders advanced polarization experiments, such as broadband Kerr or Faraday spectroscopy.

These limitations arise from intrinsic engineering constraints in strong-magnetic-field infrared platforms. To achieve high magnetic fields, solenoidal magnets are generally employed instead of split-coil designs, forcing the optical beam to propagate along the magnet bore axis over long distances. The bore aperture is typically limited to diameters of only 30-50 mm. To couple the large-diameter collimated output of a Fourier-transform infrared (FTIR) spectrometer into such a narrow bore, existing systems commonly focus the beam into a small-diameter hollow metallic light tube mounted on an insert probe, which guides the radiation toward the magnet center before final refocusing onto the sample. However, the ultra-broadband nature of FTIR radiation, spanning nearly two orders of magnitude in wavelength, renders such non-collimated transport intrinsically problematic. The converging and diverging beam experiences multiple specular reflections inside the metallic light tube. These repeated reflections not only cause significant power loss and beam divergence, but more critically, progressively scramble the polarization state. Phase differences between s- and p-polarized components upon reflection, combined with surface roughness and geometric dispersion of the light tube, lead to severe polarization degradation. As a result, linear polarization becomes partially depolarized, and circular polarization deteriorates into elliptical states with ill-defined helicity.

To mitigate polarization degradation, traditional approaches place polarization optics directly adjacent to the sample within the low-temperature, high-magnetic-field environment. This strategy relies on polarization elements compatible with cryogenic temperatures and strong magnetic fields. Such components are extremely limited in availability, often operate over narrow spectral windows, and are particularly challenging to implement for broadband infrared spectroscopy. Moreover, adjusting polarization states in this configuration requires removing the insert probe, manually rotating or replacing polarization elements, and re-cooling the system, which dramatically reduces experimental efficiency and compromising measurement reproducibility. Repeated thermal cycling, vacuum pumping, and optical realignment inevitably alter background conditions, making quantitative comparison between different polarization states unreliable.

For polarization-resolved magneto-infrared spectroscopy, a fundamentally different experimental strategy is required. Specifically, an instrument is needed that enables continuous, high-fidelity polarization control directly under in-situ sample conditions, while preserving low temperature, high magnetic fields, and a fixed optical path. In this article, in-situ polarimetry refers to the capability to continuously modulate and analyze the incident polarization on the same optical axis while the sample remains in its operational environment, without perturbing the field or cryogenic conditions.

Here, we report a collimated magneto-infrared spectroscopy system that implements in-situ polarimetry under cryogenic and high-magnetic-field conditions. The key concept is to convert the large-aperture output of FTIR spectrometer into a low-divergence beam by adding incident and exit collimation chambers to form a Kepler type optical architecture, and to guide this collimated beam through long gold-plated light tubes with a strongly reduced number of wall reflections. This collimated transport suppresses polarization scrambling and improves optical throughput, providing a robust optical backbone for polarization-resolved magneto-infrared spectrum measurements. In addition, we integrate a remotely controlled polarization module, consisting of an automated linear polarizer and a switchable Fresnel rhomb fully located outside the high-field, low-temperature region. This configuration enables continuous in-situ tuning among linear, circular, and arbitrary elliptical polarization states without thermal cycling, manual realignment, or breaking vacuum. A set of compact interchangeable focusing modules further enable the transmission and reflection measurement in both Faraday and Voigt geometries within a 50 mm magnet bore, while preserving high numerical aperture and low obscuration. Combined signal-to-noise ratio and extinction-ratio benchmarks, together with polarization-resolved measurements on representative single crystals, demonstrate that this platform overcomes the long-standing limitations of divergent-beam transport and discrete, ex-situ polarization control, and establishes a quantitative framework for polarization-resolved magneto-infrared spectroscopy in extreme environments.

## II. EXPERIMENTAL SETUP

As illustrated in Fig. 1, the complete system consists of seven main components: a vacuum FTIR spectrometer, a cryogen-free superconducting magnet, a magneto-infrared insert probe, an incident collimation chamber, an exit collimation chamber, a detector vacuum chamber, and a set of interchangeable magneto-infrared focusing modules mounted at the bottom of the insert probe and located at the center of the superconducting coil. The focusing modules are designed for various magneto-optical geometries.

The radiation source is the broadband blackbody emitter integrated into the FTIR spectrometer. After modulation by the Michelson interferometer, the beam is guided through a vacuum bellows and focused by a 90° off-axis parabolic mirror (OAP1) into the incident collimation chamber. This chamber performs several key functions: it reduces the beam size, establishes a collimated propagation section, applies chopping modulation for step-scan mode, and prepares the desired incident polarization state according to experimental requirements. Inside the incident collimation chamber, a Kepler type telescope formed by flat mirror M1 together with OAP1 and OAP2 completes the beam reduction and collimation. An aperture placed at the common focal plane of OAP1 and OAP2 is used to adjust the beam divergence. A mechanical chopper positioned near the focal region works in conjunction with step-scan modulation. The polarization modules are installed in the collimated beam segment below OAP2, enabling in-situ control of the incident polarization within the ambient environment. After beam reduction and polarization modulation, the probe beam enters the magneto-infrared insert probe. The upper end of the probe is located outside the superconducting magnet, while the lower end extends to the magnet center. The incident beam propagates to the bottom of the probe, where it is focused onto the sample surface by an interchangeable high-efficiency focusing module. By exchanging this module, the system can be configured for different magneto-optical geometries, including Faraday transmission, Voigt transmission, Faraday reflection and Voigt reflection. The reflected or transmitted signal beam from the sample is collected by the same focusing module and guided back toward the top of the probe. At the top of the probe, the emerging signal beam is redirected by OAP3 with a focal length ($f$) of 6 inches and a clear aperture of 1 inch. The beam then enters the exit collimation chamber, where it is recollimated by OAP4 and guided into the detector vacuum chamber by mirrors M2 and M3. Inside this chamber, another off-axis parabolic mirror OAP5, mounted on a motorized linear translation stage, focuses the beam onto specific infrared detectors optimized for different spectral ranges.

The FTIR spectrometer used in this system is a vacuum model, which allows a spectral resolution down to 0.3 $cm^{-1}$ and supports step-scan mode. The superconducting magnet provides a vertical static magnetic field up to 12 T. The magnet is equipped with a variable-temperature insert (VTI) with an inner bore of 50 mm, cooled by a pulse-tube refrigerator and operated in a fully cryogen-free mode. The magneto-infrared insert probe is composed of the top optical assembly, the incident and exit light tubes, and a drive rod. The top optical assembly includes two wedged diamond windows with a clear aperture of 19 mm, which provide broadband transmission from the UV to the far-infrared while maintaining the vacuum seal of the VTI. One diamond window is mounted at the top of the incident light tube, and the other at the top side of the exit light tube.

Both the incident and exit light tubes are constructed from H62 brass with polished inner walls and electroplated with a 1.5-2.0 μm gold layer to enhance reflectivity, as uniform plating over such long surfaces cannot be guaranteed. The inner diameter of each tube is 19 mm. Due to the size limitation of the electroplating bath, each tube is realized by joining two segments of 430 mm and 910.5 mm in length. To enable a monochromatic configuration, the 910.5 mm segment of the incident light tube is further divided into 888.5 mm and 22 mm sections, with the latter section at the end of the insert probe designed to be removable for installing polarization optics, as described in **Sec. III. B. 3**. The drive rod is a vacuum-compatible mechanical actuator that transmits motion to the bottom assembly without breaking the vacuum of the VTI. The driving mechanism is specially designed so that, for different focusing modules, it can selectively provide either a rotational torque (via a coupler-type interface) or a translational motion (via a screw mechanism). This flexibility allows mechanical compatibility with different sample-switching mechanisms integrated into the various focusing modules.

The detector vacuum chamber accommodates three detectors for different spectral ranges: a liquid-nitrogen-cooled HgCdTe (MCT) detector for the mid-infrared (MIR) range, a liquid-nitrogen-cooled InSb detector for the near-infrared (NIR) range, and a Si photodiode detector for the visible (VIS) range. In addition, the detector chamber is optically coupled to a 4.2 K cryogen-free silicon bolometer for far-infrared (FIR) measurements. Selection among these detectors is achieved by translating OAP5 along a motorized linear stage. The stage is equipped with multiple position sensors, enabling high repeatability in positioning and ensuring that the beam is consistently focused onto the active area of the selected detector. In practice, the operator can automatically switch detectors according to the target spectral range, enabling continuous coverage from the VIS to the FIR.

All optical paths in the system are operated either under vacuum or in a low-pressure helium environment. Fine adjustment mechanisms for optical components employ vacuum-compatible, low outgassing adjusters. In all room-temperature regions, lubricants in translation stages are replaced with vacuum greases. Stepper motors and linear stages are fully disassembled, cleaned to remove conventional lubricants, and then reassembled with vacuum grease or configured for light-load, lubricant-free operation. These measures ensure stable motion of all translation stages at the FTIR base pressure (approximately 2 mbar) without contaminating the vacuum system. All threaded holes and screws used in vacuum regions are carefully designed and checked; through holes or dedicated vacuum-compatible screws are employed to eliminate virtual leaks. Mechanical components and fasteners located in high-field regions are fabricated from non-magnetic materials such as brass, aluminum alloys, and titanium alloys, while components in weak-field regions may use weakly magnetic materials such as 316 stainless steels when appropriate.

Within this overall architecture, the incident and exit collimation chambers, together with the magneto-infrared focusing modules, constitute the key subsystems that enable in-situ polarization control. In the following, we describe the design and operating principles of the collimation system, the in-situ polarization modules, and the magneto-infrared focusing modules in detail, and present experimental and numerical characterizations of their impact on polarization preservation and signal-to-noise ratio (SNR).

## III. SETUP DESIGN AND COMPONENTS

### A. Collimation system

#### A. 1. Incident collimation chamber

As shown in Fig. 2, the incident collimation optics are housed in two vacuum chambers. The first (upper) is a 160 × 160 × 160 mm cubic chamber that contains a 4-inch-diameter, 400 mm-focal-length, 90° off-axis parabolic mirror (OAP1). OAP1 redirects the horizontally propagating collimated beam from the FTIR spectrometer downward and focuses it into the second chamber. Three external adjustment screws are used to align OAP1 with the incoming beam and to optimize its tilt and position.

The second (lower) vacuum chamber is a 204 × 210 × 484 mm rectangular enclosure mounted directly below the cube, with a vertical separation of about 5 cm between their flanges. Inside this chamber, a flat mirror M1 is mounted on a two-dimensional brass translation stage. M1 deflects the downward-propagating beam from OAP1 by 90°, sending it horizontally to the right and placing the OAP1 focal spot onto a well-defined optical axis. At this focal plane, an aperture wheel is installed. By selecting different aperture diameters, the user can trade off beam divergence and optical throughput to optimize either polarization extinction ratio or spectral SNR. Smaller apertures produce a more tightly collimated beam with lower divergence and reduced optical power, which benefits polarization extinction ratio, whereas larger apertures increase throughput at the cost of larger divergence, which favors SNR.

On the right side of the focal plane, a second off-axis parabolic mirror OAP2 is installed, with a focal length of 75 mm and a clear aperture of 1 inch. It collimates the focused beam from OAP1 and M1 and redirects it downward, sending a nearly parallel beam into the entrance of the incident light tube of the magneto-infrared insert probe. Below OAP2, we place the in-situ polarization module in the collimated beam segment. This module consists of an automated polarizer (a motorized linear polarizer that allows remote and precise angular control via external software) and a Fresnel rhomb, both mounted on vacuum-compatible motorized translation stages equipped with high-resolution encoders. Electrical connections are routed through terminal blocks and vacuum feedthroughs, enabling remote insertion/removal of polarization optics into/from the beam path and precise control of linear polarization angle under vacuum.

Compared with conventional "large divergence + direct focusing into the tube" schemes, in which a large-aperture (100 mm) OAP ($f$ = 400 mm) focuses a highly divergent beam directly onto the entrance of the light tube, the present incident collimation design converts the FTIR output into a beam with controlled and significantly reduced divergence before coupling into the tube. The two-stage OAP telescope and adjustable aperture transform the large-diameter collimated FTIR beam into a quasi-collimated beam that fills the tube with a small angular spread (e.g., ~0.382° for a 1 mm aperture of the aperture wheel in vacuum chamber). The reduced divergence has two important consequences: (i) it significantly decreases the average number of reflections on the inner wall of the light tube; (ii) it confines the most rays to strike the vacuum-metal interface at angles approaching the 90° grazing-incidence limit, leading to negligible phase shift and amplitude mismatch between s and p polarizations. Both effects are beneficial for preserving the polarization state. At the same time, fewer reflections directly improve the overall transmission efficiency of the light tube.

Both the cubic and rectangular collimation chambers are mounted on stainless-steel linear guide rails. When the magneto-infrared insert probe is prepared to be removed from the magnet, the chambers can be translated laterally by 400-500 mm along the guide rails to provide sufficient vertical clearance above the insert without disassembling any optical elements. During normal operation, the right side of the OAP1 chamber is connected to the FTIR output through a welded stainless-steel bellows, while the bottom flange is attached to the top of the insert probe. When the FTIR spectrometer is evacuated, the collimation chambers and bellows are pumped simultaneously.

A reserved port to the left of the aperture plane accommodates a mechanical chopper. Placing the chopper at this location and using lock-in detection enables step-scan measurements, providing a path for weak-signal spectroscopy in high magnetic fields and for time-resolved measurements. The walls and internal structures of both incident and exit collimation chambers are machined from 7075 aluminum alloy. Most mirror mounts and translation stages are likewise fabricated from aluminum or brass; a few components use weakly magnetic materials such as 316 stainless steels. Vacuum connections at the entrance and exit ports employ laser-welded bellows combined with sliding vacuum sealing plates. By extending or compressing the bellows and translating the sealing plates, one can finely adjust the relative position between the FTIR output and the collimation chamber while keeping the relative alignment of OAP2, the magnet, and the insert probe fixed. This mechanical arrangement greatly simplifies global optical alignment and long-term maintenance.

**A. 2. Exit collimation chamber**

As shown in Fig. 3, the exit collimation optics are also divided into two vacuum chambers. The left chamber houses an OAP4 and M2. OAP4 recollimated and reflects beam downward onto M2. M2 redirects the beam horizontally to the right into a room-temperature light tube with polished and gold-plated inner walls. External adjustment screws on OAP4 and M2 allow fine tuning of their tilt and position, so that the output beam divergence and propagation direction can be optimized. In the aligned configuration, the collimated beam experiences at most a single reflection in the room-temperature light tube, minimizing reflection losses and preserving polarization fidelity. After traversing the first room-temperature light tube, the beam enters vacuum chamber, which contains an adjustable mirror M3. M3 reflects the beam downward into a second gold-plated light tube that guides the beam into the detector vacuum chamber.

The light tube connecting the two exit chambers is constructed from H62 brass with a polished inner surface and a gold plating. The composite structure has an outer diameter of 32 mm, an inner diameter of 28 mm, and a total length of 900 mm. Due to the difficulty of uniformly polishing and gold-plating a long and narrow tube, and because thermal contraction is negligible at room temperature, we adopt a “short-segment + outer sleeve” design. Specifically, three 200 mm and two 150 mm brass segments (total length 900 mm) are inserted into a smooth, unplated outer brass sleeve with an inner diameter of 32 mm, outer diameter of 38 mm, and length of 650 mm. The joints between the short inner segments are fully covered by the outer tube. Before assembly, the ends of the short tubes are coated with a thick layer of vacuum epoxy. The segments are then slowly inserted into the outer tube so that epoxy fills the gaps between inner and outer tubes. After curing, the composite structure becomes vacuum-tight and can be treated as a single 900 mm light tube with a continuous gold-plated inner wall. The end faces of the composite tube are machined on a lathe to reduce any misalignment at the joints to a negligible level.

The second light tube between M3 and the detector chamber has an inner diameter of 36 mm, outer diameter of 38 mm, and length of 480 mm. It is fabricated using the same polishing, plating, and epoxy-bonded composite technique.

**A. 3. Collimation effect**

To quantitatively evaluate the advantage of the present collimation scheme over a conventional “large-divergence + direct focusing into the tube” configuration, we analyze both the number of inner wall reflections experienced by rays inside the incident light tube and the resulting polarization preservation.

We consider a cylindrical light tube with length $L = 1340$ mm and inner diameter $2R = 19$ mm as in the setup. A geometric factor can be defined as $k = \frac{L}{2R} \approx 70.526$. For a given ray entering the tube, its trajectory is described in a plane containing the tube axis. The ray starts at a radial position $x_0$, with $0 \leq x_0 \leq R$, and propagates at a small angle $\alpha$ relative to the tube axis, with $|\alpha| \ll 1$ rad. We define the radial slope as $t = \tan(\alpha)$. In an “unfolded” representation, the radial coordinate evolves approximately as $x_{\text{unfold}}(z) = x_0 + s \cdot t \cdot z$, where $z$ is the axial coordinate along the tube and $s = \pm 1$ denotes the initial radial direction (outward or inward). Specular reflection at the cylindrical wall $x = \pm R$ corresponds to a crossing of these boundaries in the unfolded coordinate. The axial position of the first wall reflection is given by $z_1 = \frac{x_{\text{boundary}} - x_0}{s \cdot t} = \frac{R - \frac{x_0}{s}}{t}$, where $x_{\text{boundary}} = s \cdot R$ is the closest tube wall along the ray direction. If $z_1 \leq 0$ or $z_1 \geq L$, the ray exits the tube without touching the wall and the total number of reflections is $N = 0$. If $0 < z_1 < L$, the ray undergoes at least one reflection. In an ideal cylindrical tube, each subsequent reflection corresponds to a radial traversal of $2R$ and therefore an axial spacing $\Delta z = \frac{2R}{t}$. The remaining axial length after the first reflection is $z_{\text{rem}} = L - z_1$. The additional number of reflections is then $N_{\text{extra}} = \text{floor}\left(\frac{z_{\text{rem}}}{\Delta z}\right) = \text{floor}\left(\frac{(L - z_1)t}{2R}\right)$. The floor function returns the greatest integer less than or equal to the argument. Thus, for $0 < z_1 < L$, the total number of reflections is $N = 1 + \text{floor}\left(\frac{(L - z_1)t}{2R}\right)$, whereas for $z_1 \leq 0$ or $z_1 \geq L$, $N = 0$.

In a conventional design commonly used in magneto-infrared spectroscopy, a 100 mm aperture off-axis parabolic mirror with focal length $f = 400$ mm directly focuses the FTIR beam into the entrance of the light tube. The focal spot approximately fills the tube diameter, and the chief ray is aligned with the tube axis. In this case, the maximum half-angle of rays entering the tube is $\alpha_{\text{max}} = \arctan\left(\frac{50}{400}\right) \approx 7.125°$ corresponding to $t_{\text{max}} = \tan(\alpha_{\text{max}}) = 0.125$. Assuming a uniform intensity distribution over the OAP pupil, the geometric relation gives $r = f \cdot \tan(\alpha)$ and $\frac{\mathrm{d}r}{\mathrm{d}\alpha} = f \cdot \sec^2(\alpha)$. The corresponding angular probability density is therefore $p(\alpha)\mathrm{d}\alpha \propto r\mathrm{d}r \propto \tan(\alpha) \cdot \sec^2(\alpha)\mathrm{d}\alpha$. Substituting $t = \tan(\alpha)$ and $\mathrm{d}t = \sec^2(\alpha)\mathrm{d}\alpha$, the ray-angle distribution expressed in terms of $t = \tan(\alpha)$ simplifies to $p(t) \propto t$ within the interval $t \in [0, t_{\text{max}}]$. Considering rays entering near the tube axis ($x_0 \approx 0$), the first reflection occurs at $z_1 \approx \frac{R}{t}$, and subsequent reflections are spaced by approximately $\Delta z \approx \frac{2R}{t}$. From this simplified on-axis model, analytic binning conditions for the reflection number can be obtained. No reflection ($N = 0$) requires $z_1 > L$, i.e., $t < t_0^+ = \frac{R}{L}$. For $N = n \geq 1$, the condition $z_n \leq L < z_{n+1}$, with $z_n \approx \frac{(2n-1)R}{t}$, gives an angular interval $t \in [t_n^-, t_n^+)$, where $t_n^- = \frac{(2n-1)R}{L}$ and $t_n^+ = \frac{(2n+1)R}{L}$. After truncation to $t \in [0, t_{\text{max}}]$, the probability of each reflection number $P(N = n)$ can be evaluated analytically. Using realistic parameters, this conventional scheme yields a substantial fraction of rays undergoing multiple reflections, with maximum reflection numbers reaching $N = 9$.

In the present collimated scheme, the FTIR beam is first focused by OAP1 onto an adjustable aperture, which defines the effective numerical aperture and divergence. The beam is then recollimated by OAP2 before entering the incident light tube. To determine the divergence angles for different aperture diameters, the light tube is removed, and the far-field beam profiles are measured directly; the measured beam diameters at known distances are used to extract the corresponding angular spreads. These experimentally determined divergences are then used as inputs for the geometric model. For each aperture diameter $D$, we perform a Monte Carlo simulation in which typically $6 \times 10^4$ rays are launched. The initial radial positions $x_0$ are uniformly distributed over the illuminated area at the tube entrance $\left(0 \leq x_0 \leq r_{\max}(D)\right)$ and the initial radial direction $s = \pm 1$ is chosen with equal probability. For each ray, the total reflection number $N$ inside the tube is calculated using the geometric rules described above, yielding a statistical distribution of reflection counts for each aperture setting.

The simulation demonstrates that the present collimation scheme strongly suppresses higher-order reflections (Table 1). For typical aperture sizes, the majority of rays undergo at most zero or one reflection. For example, with a 1.0 mm aperture, approximately 25 percent of rays experience a single reflection, while higher-order $(N \geq 2)$ reflections are essentially eliminated. This is in sharp contrast to the conventional direct-focusing configuration, where multi-reflection trajectories are common.

To relate these geometric results to experimental polarization performance, we measure the linear polarization extinction ratio of the complete system. A rotatable wire grid polarizer is positioned immediately before the entrance of the incident light tube, and a fixed infrared analyzer is placed at the sample position. The transmitted intensity is recorded by the MCT or InSb detector in the detector vacuum chamber. Both polarizers have a nominal extinction ratio of approximately 250:1. With a 1 mm aperture in the incident collimation chamber, the measured system extinction ratio reaches up to 40:1. Increasing the aperture to 2 mm reduces the extinction ratio to approximately 15:1. For aperture diameters between 3 and 7 mm, the extinction ratio approaches a plateau of roughly 7:1. These results are consistent with the simulated reflection statistics. Even in the 1 mm aperture case, about 25 percent of rays undergo one reflection on the gold-plated inner wall of the tube. At near-grazing incidence, such reflections still introduce a finite additional phase difference between s- and p-polarized components. Together with residual surface roughness and azimuthal dispersion along the tube circumference, this leads to partial degradation of polarization purity. As the aperture increases from 1 mm to 2-3 mm, most rays experience at least one reflection, and the system enters a regime in which the extinction ratio decreases from about 15:1 toward a stable value near 7:1, following the same trend predicted by the geometric model.

In summary, the combined incident and exit collimation system converts the large-diameter FTIR output into a low-divergence beam and limits the number of wall reflections during propagation through the light tubes. By strongly suppressing multi-reflection procedure, the collimation scheme improves both optical throughput and polarization preservation, providing a foundation for high-fidelity in-situ polarimetry in strong-magnetic-field magneto-infrared spectroscopy.

**B. In-situ polarization configuration**

To enable polarization-resolved spectroscopy under in-situ conditions, a configurable polarization module is implemented inside the incident collimation chamber (Fig. 4).

The module consists of an automated polarizer and a Fresnel rhomb (or equivalent phase-retarding prism), each of which can be independently inserted into or removed from the beam path by motorized linear translation stages. By coordinating the motion of these elements, the system can be switched among three operating modes: a non-polarization mode, an in-situ linear polarization mode, and an in-situ circular polarization mode. In addition, the magneto-infrared insert probe retains the capability of accommodating polarization optics directly at low temperature and high magnetic field, providing design redundancy for specialized experiments.

All polarization-state switching and angle control are performed via external control software, without breaking vacuum or disturbing the sample temperature, magnetic field, or optical alignment. This configuration allows continuous and reproducible polarization scans while the sample remains in its operational environment.

**B. 1. In-situ linear polarization configuration**

In the in-situ linear polarization configuration (Fig. 4b), the automated polarizer is inserted into the collimated beam section located beneath OAP2 in the incident collimation chamber. The black-body light source is unpolarized. After beam reduction and collimation by OAP1 and OAP2, the incident beam is converted into a linearly polarized state with a chosen polarization angle. The polarized beam then propagates through the incident light tube, the magneto-infrared focusing module, and finally impinges on the sample. During experiments, the aperture wheel in the incident collimation chamber can be used to select an appropriate aperture diameter. Smaller apertures produce lower beam divergence and higher polarization purity, which are advantageous for polarization-sensitive measurements. Larger apertures increase optical throughput and are preferred when the signal level is weak and higher SNR is required.

The rotation angle $\theta$ of the automated polarizer can be specified by the user as a series of discrete values (for example, in steps of 5° or 10°). At each polarization angle, the system automatically executes a complete magneto-infrared spectroscopy measurement sequence, while keeping all other parameters, such as sample position, temperature, and magnetic field unchanged. In this way, a set of spectra corresponding to different linear polarization orientations is obtained under identical experimental conditions. Compared with traditional approaches that rely on manual rotation of polarizers and require removal and reinsertion of the probe followed by re-cooling, the in-situ linear polarization scheme greatly improves experimental efficiency and ensures quantitative comparability between different polarization angles.

**B. 2. In-situ circular polarization configuration**

For circular-polarization-resolved magneto-infrared spectroscopy (Fig. 4c), the system operates by extending the linear polarization configuration described above. A Fresnel rhomb is inserted into the beam path below the automated polarizer via a motorized translation stage, forming a broadband achromatic circular polarizer. Unlike conventional wave plates, a Fresnel rhomb generates a relative phase delay between s- and p-polarized components through multiple total internal reflections inside the rhomb. The resulting phase retardation exhibits only weak wavelength dependence over the transmission band of the substrate material, making the Fresnel rhomb particularly suitable for broadband magneto-infrared applications. In this work, a ZnSe Fresnel rhomb combined with a Thallium Bromoiodide (KRS-5) automated wire-grid polarizer is used as a representative implementation for the MIR range. For extension to other spectral regions, alternative materials such as high-resistivity silicon, calcium fluoride, sapphire, quartz, or other infrared materials can be employed.

In this configuration, the incident beam is first linearly polarized, with the electric-field vector oriented at $\pm 45°$ with respect to the principal axis of the Fresnel rhomb. After passing through the rhomb, the s- and p-components acquire a relative phase shift of approximately $\pi/2$, resulting in circularly polarized output. Compared with linear polarization, circular polarization is less sensitive to azimuthal dispersion and surface-normal variations of the inner walls of the light tube. As a result, the polarization state is generally better preserved under multiple near-grazing reflections. In practice, this allows the use of a larger aperture than in the linear polarization mode, thereby improving spectral SNR.

Because the insertion of the Fresnel rhomb slightly alters the output beam position from the incident collimation chamber, a corresponding mechanical compensation is implemented. The entire OAP1 chamber can be translated along the stage, and the vacuum bellows connecting it to the FTIR spectrometer can be extended or compressed as needed. This adjustment allows the modified beam to be realigned with the entrance of the magneto-infrared insert probe without changing the position of the chamber exit port. By rotating the linear polarizer angle, the handedness of the circular polarization can be switched between left- and right-circular states, and intermediate elliptical polarizations can be accessed continuously. All of these operations are carried out without disturbing the low-temperature or high-magnetic-field environment, enabling in-situ circular-polarization-resolved measurements.

**B. 3. Polarization measurement of monochromatic light**

In specific experiments, more stringent polarization control is required at a specific wavelength or over a very narrow spectral range. In this case, a monochromatic polarization configuration can be employed (Fig. 4d). Polarization optics compatible with cryogenic temperatures and high magnetic fields can be modularly installed at the bottom of the incident light tube, close to the sample position. As mentioned earlier in **Sec. II,** the incident light tube includes a removable section near the focusing module, allowing polarization optics to be installed at the end of the insert rod. In this mode, the polarization module inside the incident collimation chamber is set to the non-polarization configuration by retracting both the automated polarizer and the Fresnel rhomb from the beam path. The incident beam propagates through the light tube is then converted into linearly or circularly polarized light by the low-temperature polarization elements immediately before entering the focusing optics. Because the polarization state is generated in close proximity to the sample, this configuration minimizes the influence of polarization degradation caused by reflections within the light tube. As a result, improved polarization fidelity can be achieved at the chosen wavelength. However, this monochromatic mode does not allow in-situ adjustment of the polarization state during measurement, and the limited space available within the high-field region precludes the use of broadband achromatic elements such as Fresnel rhombs. Consequently, circular polarization in this mode is typically restricted to single-wavelength operation. This configuration therefore serves as a complementary option for specialized measurements requiring maximum polarization purity at a specific photon energy, rather than for broadband polarization scans.

### C. Magneto-infrared focusing modules

After polarization modulation and collimated transport through the incident light tube, the beam reaches the bottom of the magnet bore, where a set of changeable magneto-infrared focusing modules performs three functions: focusing onto the sample, enabling the desired Faraday or Voigt geometry, and collecting and re-directing the transmitted or reflected signal light back into the exit light tube.

In a solenoidal superconducting magnet, increasing the sample-space diameter rapidly enlarges the coil radius and the associated hoop stress, pushing the mechanical limits of the superconducting wire and support structure. In addition, constraints on ampere-turn density, cooling channels, and mechanical reinforcement limit the available bore diameter. As a result, the usable sample-space diameter is typically restricted to 30-50 mm. The present system uses a 50 mm inner-diameter VTI, which is relatively generous by high-field standards, but still imposes stringent requirements on the design of magneto-infrared focusing modules. Within this limited cylindrical volume, the modules must provide high numerical-aperture (NA) focusing, low obscuration, and good polarization preservation, while also accommodating multi-sample holders and mechanical actuation.

To meet these requirements, we developed four modular focusing modules (Fig. 5): Faraday transmission, Voigt transmission, Faraday reflection, and Voigt reflection. All four modules share a common mechanical interface to the bottom of the insert probe, and each integrates its own sample holder and internal actuation mechanism driven by the external drive rod. This modular approach allows rapid switching between geometries by exchanging the entire focusing module, without changing the upstream optics.

### C. 1. Faraday transmission focusing module

In the Faraday transmission configuration (Figs. 5(a)-5(c)), the beam propagation direction is parallel to the magnetic field. The incident collimated beam enters the Faraday transmission module coaxially through the incident light tube. Inside the module, an OAP first reflects and focuses the beam, bending it toward the sample. A right-angle prism then redirects the beam again so that, at the sample position, the propagation direction is aligned with the magnetic field axis, satisfying the Faraday geometry. After passing through the sample, the transmitted beam is collected by a second OAP-prism pair arranged symmetrically with respect to the sample plane and is recollimated into the exit light tube.

Because the entire optical module, including mirrors, prisms, sample gear, and mechanical supports, must fit within the 50 mm (effectively ≈ 48 mm) diameter cylindrical space of the VTI, the optical layout cannot be realized as a simple planar, rotationally symmetric system. Instead, the two OAPs and two right-angle prisms are arranged in a compact three-dimensional, non-axisymmetric geometry with staggered components. The samples are mounted on a "sample gear" positioned between the prisms and laterally offset from the symmetry plane to minimize obscuration.

The Faraday transmission module is designed to accommodate up to nine samples in a single cooldown. The drive rod at the top of the insert probe is mechanically coupled to an internal leadscrew, which rotates a transmission gear at the bottom of the module. Rotational motion is transmitted via a triangular form-fit coupling between the drive rod and triangular boss of the leadscrew. This transmission gear meshes with the sample gear, causing the sample gear to rotate and sequentially bring different samples into the optical path. The sample gear contains nine evenly spaced apertures with a diameter of 2.2 mm. The rear side of each aperture is machined into a conical surface to minimize obstruction of the transmitted beam. Samples are mounted over these apertures and clamped from the front side by a sample cover flap, which is also shaped with a conical opening to avoid shading the beam. For convenient sample loading, the sample gear is mounted on a pivot shaft that allows it to be swung out of the optical path (Fig. 6). After samples are installed, the gear is swung back into the working position. Inside the module, a spring clip presses on the rotation shaft of the sample gear to eliminate fit clearance and maintain the sample gear in a stable and reproducible position during spectral measurements.

**C. 2. Voigt transmission focusing module**

In the Voigt geometry (Figs. 5(d)-5(f)), the beam propagation direction is perpendicular to the magnetic field. In the Voigt transmission module, the collimated beam emerging from the incident light tube is reflected and focused by an OAP onto the sample. A second OAP, placed symmetrically with a shared focal point, collects the transmitted light and redirects it into the exit light tube. This compact two-OAP confocal arrangement ensures that the beam direction at the sample is strictly perpendicular to the magnetic field and to the sample surface, while simultaneously minimizing the spot size, reducing obscuration, and preserving polarization. The original concept of this two-OAP Voigt geometry was proposed by Mykhaylo Ozerov at the National High Magnetic Field Laboratory[32].

As in the Faraday transmission module, multi-sample capability is incorporated to increase throughput. The Voigt transmission module can host up to eight samples per cooldown. The samples are mounted on a vertically movable sample holder with eight transmission apertures arranged along the direction of motion. Unlike the Faraday configuration, where the samples are selected by rotating a gear in the horizontal plane, the Voigt module employs a drive rod coupled to an internal leadscrew via a threaded interface. In this configuration, the internal thread of the drive rod directly engages with the thread of the leadscrew. Rotation of the drive rod by an external motor causes the leadscrew to translate the sample holder vertically, sequentially positioning each sample at the common focal point of the two OAPs. The sample holder and sample cover flap are machined with conical openings to avoid blocking the transmitted beam. To prevent accidental disengagement of the threaded interface between the drive rod and the leadscrew, the module incorporates an elastic upper limit mechanism consisting of a spring and piston assembly, and an elastic lower limit mechanism comprising a spring strip and an adjustable screw. These mechanisms provide compliant end stops that limit the travel range and maintain thread engagement. Furthermore, to ensure high positional stability during sample switching, a dovetail slider with double-sided spring clips is used to eliminate backlash while preserving smooth motion of the sample holder.

**C. 3. Reflection focusing modules**

The system also supports Faraday and Voigt configurations in reflection geometry, which is essential for opaque samples or for magneto-infrared measurements. In the Faraday reflection module (Figs. 5(g)-5(i)), a single on-axis parabolic mirror serves as both focusing and collecting optic. Emerging from the light tube, the incident beam traverses the central aperture of the ring-shaped sample stage and reaches the on-axis parabolic mirror located at the bottom of the module. The mirror focuses the beam onto the sample located near its focal plane. After reflection from the sample, the beam is recollimated by the same parabolic mirror and exits back. To suppress stray light at large angles of incidence, a coaxial graphite-coated light baffle is installed above the mirror. To accommodate multiple samples while preserving the central beam path, the sample stage is realized as a removable ring gear. Up to eight samples can be mounted around the ring. The entire sample stage can be taken out of the module for sample mounting and then reinserted. Sample selection is achieved via a gear-driven mechanism similar to that used in the Faraday transmission module, with the drive rod and internal transmission gear rotating the ring gear to bring each sample in sequence into the focused beam. In the Voigt reflection module (Figs. 5(j)-5(l)), a pair of OAPs arranged in a confocal send–receive geometry is used. The incident beam from the light tube is reflected and focused by the first OAP onto the sample at the common focal point of the two OAPs. The reflected beam is then collected by the second OAP and redirected toward the exit light tube. A graphite-coated light baffle is placed between the two OAPs to block stray reflections and suppress background. The geometrically symmetric design of the two reflection modules and careful control of obscuration is favorable for preserving the polarization state of the beam in both Faraday and Voigt reflection geometries.

Taken together, these modular focusing modules provide a unified and flexible platform for magneto-infrared measurements in all four major geometries (Faraday/Voigt, transmission/reflection). Their compact, mechanically robust designs, in conjunction with the external collimation and polarization subsystems, establish stable optical conditions for high-fidelity in-situ polarimetry under strong magnetic fields and at cryogenic temperatures.

## IV. PERFORMANCE

### A. Signal-to-noise ratio

The SNR of a magneto-infrared spectroscopy system determines the minimum field-induced optical response that can be reliably resolved and is therefore one of the most critical performance metrics. To evaluate the effectiveness of the collimated optical transport and the overall SNR of the system, we characterize the noise level under polarization-free conditions in different spectral ranges.

As shown in Fig. 7, we perform infrared reflectivity measurements on a 100 nm thick Au film and on a single-crystalline Sb sample under identical temperature and magnetic-field conditions. Two reflection spectra are collected for each sample under the same experimental settings, yielding $R_{Au}$ and $R_{Sb}$. The noise spectrum is obtained by taking the ratio of two successive measurements, $R_{Au}/R_{Au}$ and $R_{Sb}/R_{Sb}$, respectively. In the ideal case (i.e., without noise, source light fluctuation and other disturbances), this ratio should equal unity over the entire spectral range.

All measurements are performed with an acquisition time of 1 min per spectrum. The resulting noise spectra are shown in Fig. 7. In both cases, the fluctuation amplitude is well below 0.1% over most of the spectral range. To quantify the noise level, we evaluate the root-mean-square (RMS) noise within sliding spectral window of width 200 $cm^{-1}$. The average RMS noise is found to be 0.0082%, with a minimum value reaching as low as 0.0033%. These results demonstrate that the collimated optical design and vacuum optical environment enable a very low noise floor, which is essential for detecting weak magneto-infrared signals.

**B. Extinction ratio in linear polarization mode**

The polarization performance in linear polarization mode is characterized by the extinction ratio, measured using the polarizer-analyzer method described in **Sec. III. A. 3**. An automated polarizer is inserted into the polarization module inside the incident collimation chamber, while an analyzer is placed at the sample position. By continuously rotating the incident polarizer, Fourier-domain intensity spectra are recorded at both the intended parallel and cross polarization configurations.

Fig. 8 shows the measured extinction ratio in two representative spectral ranges for an aperture diameter of 1 mm. Over most of the spectrum, the extinction ratio exceeds 20:1 and reaches values as high as 40:1 at optimal frequencies. A pronounced reduction of both signal intensity and extinction ratio is observed near the edges of the spectral window, which is primarily attributed to the cutoff characteristics of the detector and the beamsplitter. The absorption feature near 2000 $cm^{-1}$ originates from the diamond windows used as vacuum seals.

We further perform systematic measurements of the overall extinction ratio for different aperture diameters in the incident collimation chamber. The results are fully consistent with the geometric and Monte Carlo analyses presented in **Sec. III. A. 3**. With a 1 mm aperture, the extinction ratio reaches approximately 40:1. Increasing the aperture to 2 mm reduces the extinction ratio to about 15:1, while apertures in the range of 3-7 mm yield a stable extinction ratio of approximately 7:1. These values are sufficient for the majority of polarization-resolved magneto-infrared experiments. For experiments requiring higher polarization purity, a smaller aperture can be selected at the expense of optical throughput.

**C. In-situ linear polarimetry on arsenic in Voigt reflection**

The in-situ linear polarization capability of the system is demonstrated by polarization-resolved reflectivity measurements on single crystal arsenic in the Voigt reflection geometry, as shown in Fig. 9. In this configuration, the linear polarization angle of the incident infrared beam can be continuously varied under in-situ conditions.

In the experiment, the sample, temperature, and magnetic field are kept unchanged, while the linear polarization angle $\theta$ is rotated from 0° to 360° with a step size of 5°. Measurements are performed at 0 T and 12 T, and the relative reflectivity spectra are computed as $R_{12\,\mathrm{T}}/R_{0\,\mathrm{T}}(\theta)$ for each polarization angle.

The resulting spectrum exhibits pronounced polarization-dependent features under magnetic field. Spectral signatures, possibly associated with Landau-level-related excitations or other field-induced collective modes, are strongly enhanced near polarization angles of approximately 90° and 270°. The overall angular dependence displays a clear $C_2$ rotational symmetry, fully consistent with the expected behavior for linear polarization in the Voigt geometry. This experiment verifies that the highly collimated magneto-infrared system enables reliable in-situ linear polarimetry with continuous angular control and high reproducibility.

**D. In-situ circular polarimetry on MBT in Faraday transmission**

To demonstrate in-situ circular polarization capability, we perform magneto-infrared transmission measurements on a $Mn(Bi,Sb)_2Te_4$ (MBT) sample in the Faraday transmission geometry using the in-situ circular polarization module. Left- and right-circular polarization states are realized by adjusting the relative orientation between the automated polarizer and the Fresnel rhomb, and the corresponding transmission spectra are recorded.

In the measurement, the polarization angle is scanned from 0° to 360° with a step size of 3°. Spectra are acquired at both 0 T and 12 T, and the relative transmission is plotted as $T_{12\,\mathrm{T}}/T_{0\,\mathrm{T}}(\theta)$, as shown in Fig. 10. At polarization angles corresponding to integer multiples of 90°, the incident light is circularly polarized, while intermediate angles correspond to elliptically polarized states.

The measured spectrum exhibits a clear $C_2$ periodicity as a function of polarization angle, consistent with the switching between different circular polarization states. Owing to the achromatic nature of the Fresnel rhomb, this behavior persists over a broad infrared spectral range. In the antiferromagnetic phase, MBT hosts spin-degenerate insulating bands, and optical absorption without polarization discrimination corresponds to a band-edge transition. Upon applying a magnetic field and entering the ferromagnetic phase, spin degeneracy is lifted and the band structure is modified, leading to a field-induced shift of the band-edge absorption.

Under circularly polarized excitation, the antiferromagnetic phase at zero field respects the combined $\mathcal{PT}$ symmetry, resulting in identical absorption for both circular polarizations. In contrast, in the ferromagnetic phase, $\mathcal{PT}$ symmetry is broken and the band topology is modified, giving rise to different absorption strengths for the two circular polarizations. Previous studies have attributed magnetic circular dichroism in such systems to Berry curvature emerging after field-induced topological transitions, often accompanied by anomalous Hall responses[45]. The present experiment demonstrates that the developed system enables in-situ circular-polarization-resolved magneto-infrared spectroscopy over a broad spectral range.

## V. OUTLOOK

Future development of the system includes integration with infrared microscope for spatially resolved magneto-infrared imaging, extension to pulsed magnetic fields that are beneficial for time-resolved studies, and femtosecond laser-based pump-probe measurements for investigating magneto-optical response dynamics. At this time, performing pump-probe time-resolved measurements over a broadband infrared spectral range remains challenging and requires further efforts. Alternatively, in the present instrument, time-resolved infrared measurements can be realized using the step-scan mode of the spectrometer, in which the interferometer moveable mirror is held at fixed positions to record time-dependent signals and reconstruct the temporal evolution of spectra. Beyond condensed-matter physics, the system also holds promise for applications in other fields, including the characterization of organic polymers in chemistry, infrared-based cancer imaging and diagnosis in biology and medicine, and magneto-optical filtering for measurements of solar magnetic and velocity fields in astronomy.

## VI. CONCLUSION

In conclusion, we have developed a collimated in-situ polarimetry system for magneto-infrared spectroscopy. Incident and exit collimation chambers form a stable Kepler type collimated optical path that suppresses beam divergence and minimizes reflections inside the light tubes. On this basis, a remotely controlled polarization module placed outside the cryogenic high magnetic field region enables continuous in-situ tuning of linear and circular polarization states. The improved collimation enhances polarization fidelity, and a set of magneto-infrared focusing modules supports Faraday and Voigt configurations in both transmission and reflection. Together, these components establish a versatile and polarization-resolved platform for magneto-infrared studies under cryogenic and high-magnetic-field conditions. Performance characterization shows an extinction ratio reaching approximately 40:1 in linear polarization mode and a minimum RMS noise of 0.0033%. Experimental validation includes broadband in-situ continuous linear polarization measurements in Voigt reflection configuration and broadband in-situ continuous circular polarization measurements in Faraday transmission configuration at 1.7 K and 12 T. Both experiments exhibit distinct polarization-dependent magneto-infrared signatures, marking the first realization of continuous polarization-resolved measurements in magneto-infrared spectroscopy under these conditions. This work provides a reliable experimental platform for polarization-resolved spectroscopy, opening new possibilities for investigating selective optical responses in quantum materials under extreme environments.

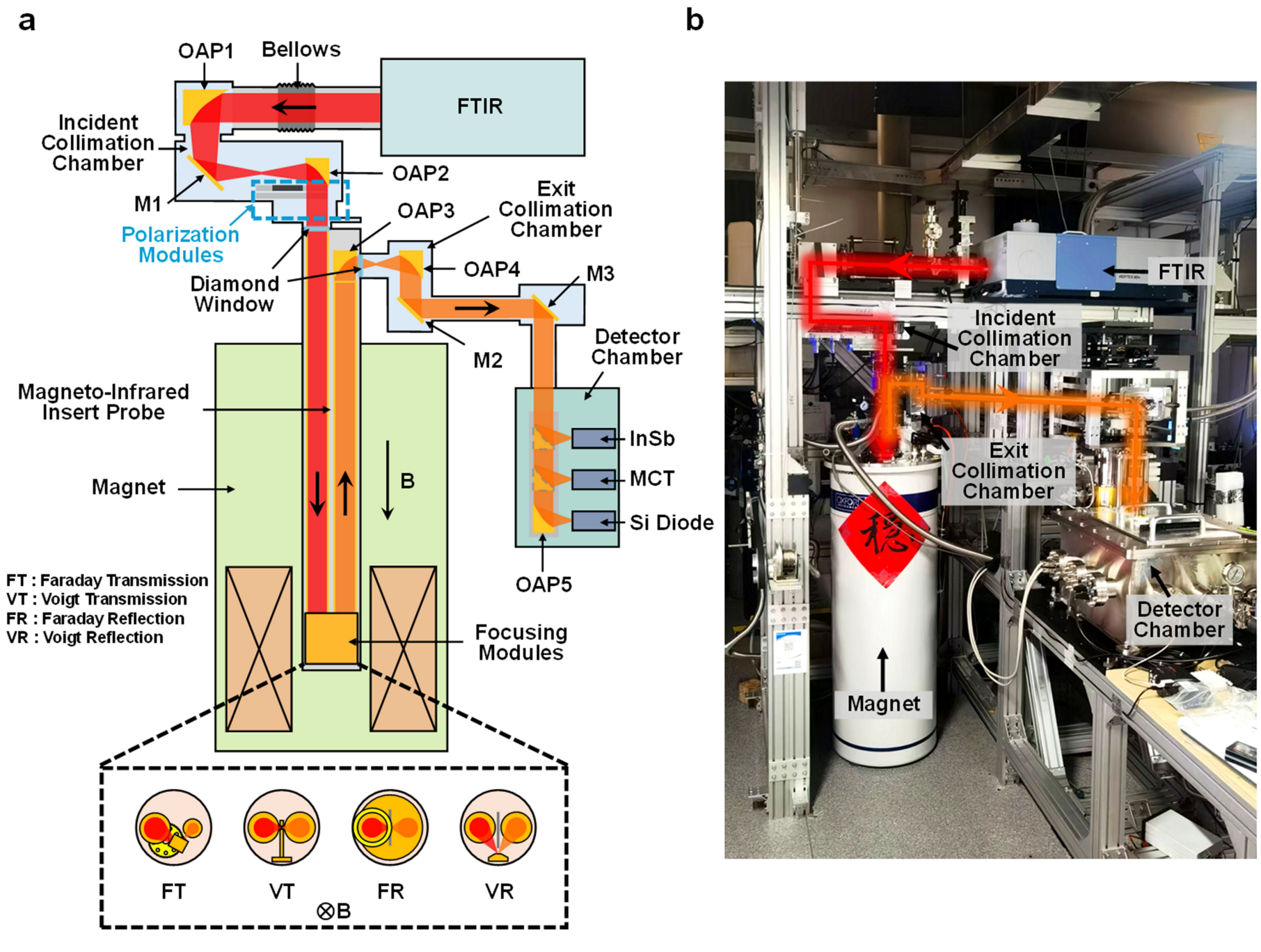

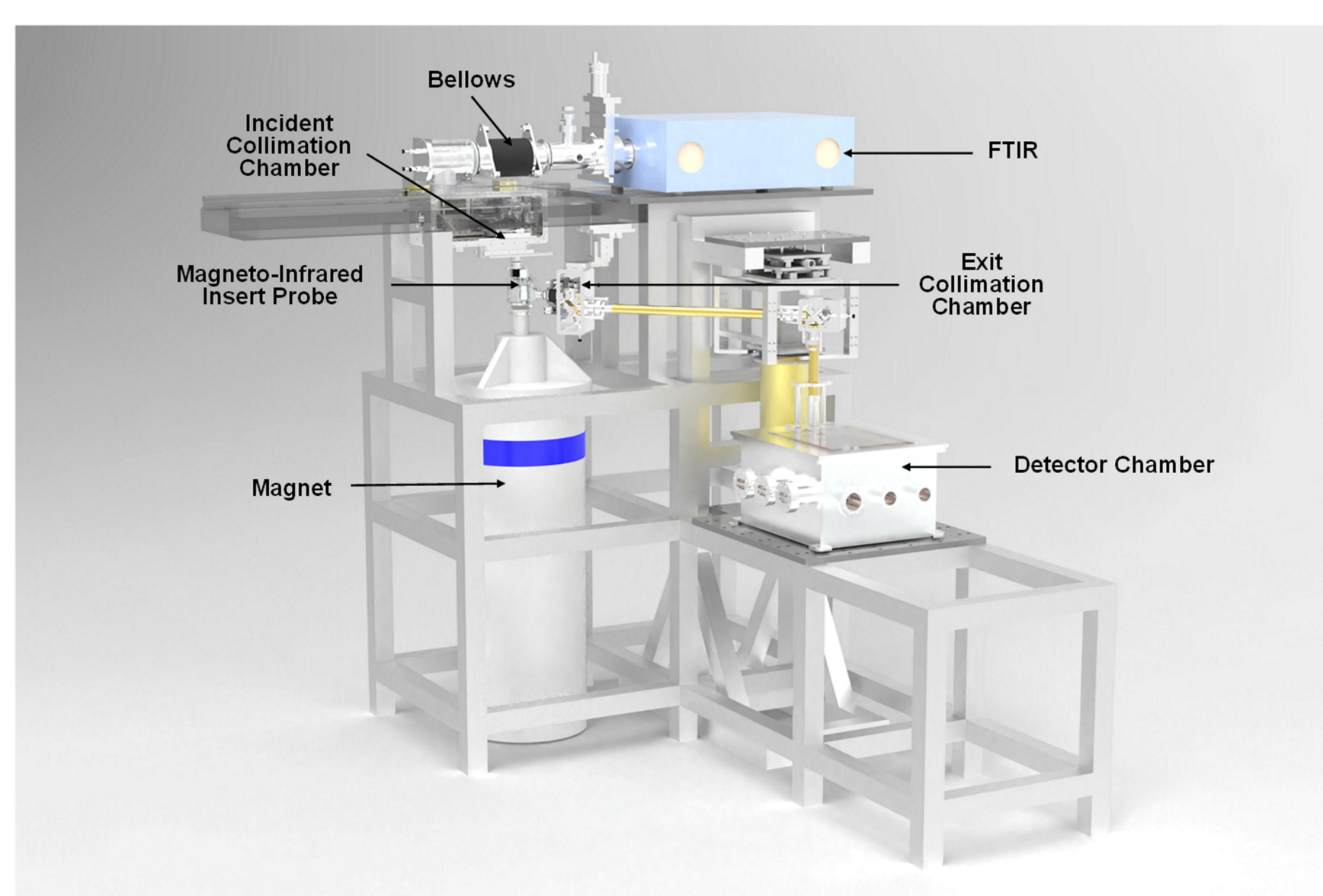

**Fig. 1** Schematic plot and photos of the collimated magneto-infrared spectroscopy system. The broadband and modulated radiation from the FTIR spectrometer is guided into the incident collimation chamber, where it is focused by OAP1, redirected by flat mirror M1, and subsequently recollimated by OAP2 before being coupled into the magneto-infrared insert probe. The probe guides the beam to the sample located at the magnetic field center with an interchangeable magneto-infrared focusing module. The reflected or transmitted signal light from the sample is collected back through the probe, redirected by OAP3 into the exit collimation chamber, recollimated by OAP4, and steered by flat mirrors M2 and M3 into the detector vacuum chamber. Finally, the signal is focused onto the selected detector by OAP5.

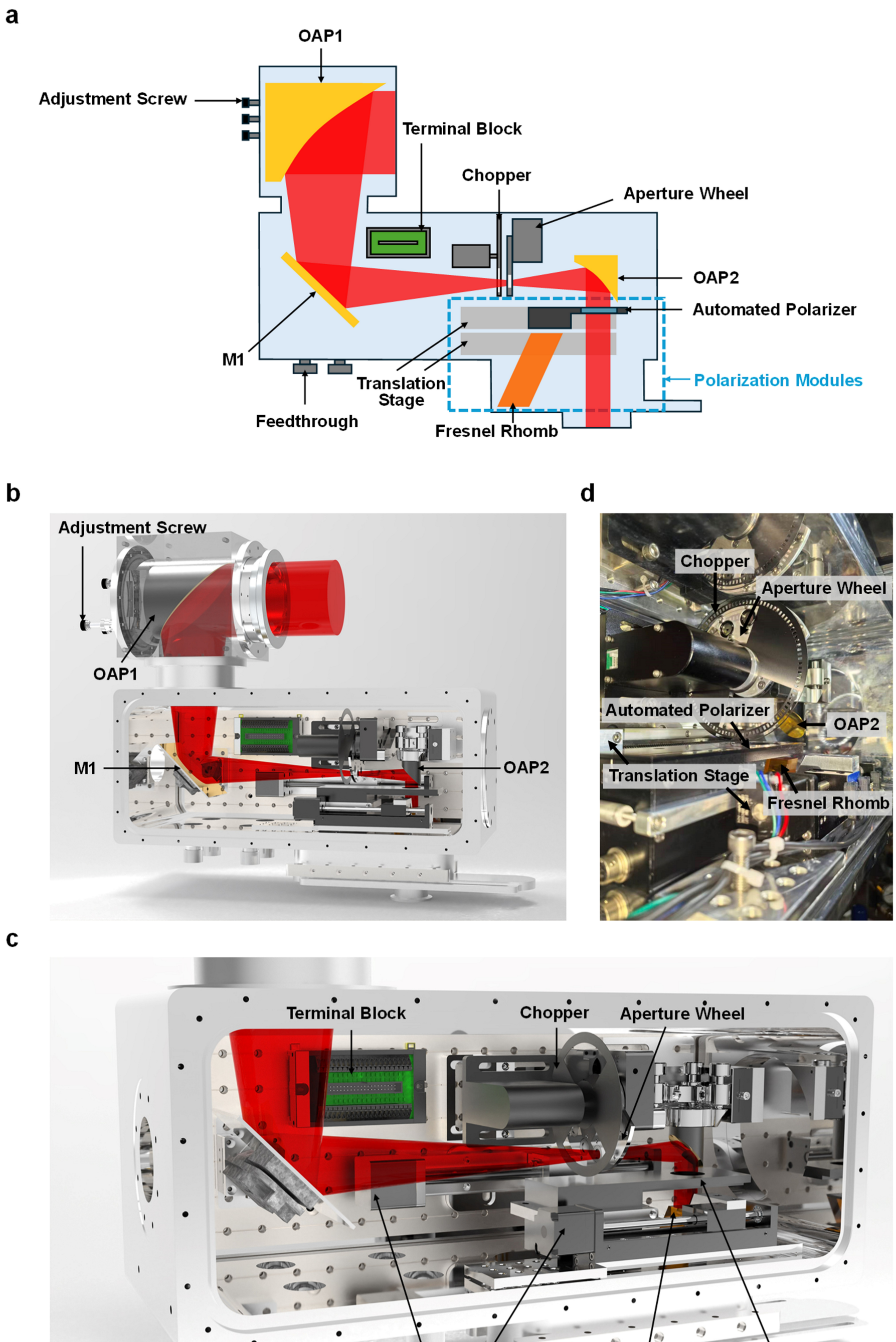

**Fig. 2** Design and photo of the incident collimation chamber. This chamber performs beam reduction and collimation before the light enters the magneto-infrared insert probe. A Kepler type telescope formed by OAP1, M1, and OAP2 converts the large-diameter FTIR spectrometer output into a quasi-collimated beam with adjustable divergence, controlled by an aperture wheel. The in-situ polarization module, consisting of an automated polarizer and a switchable Fresnel rhomb installed on motorized translation stages, enables in-situ polarization modulation entirely outside the cryogenic high-field region.

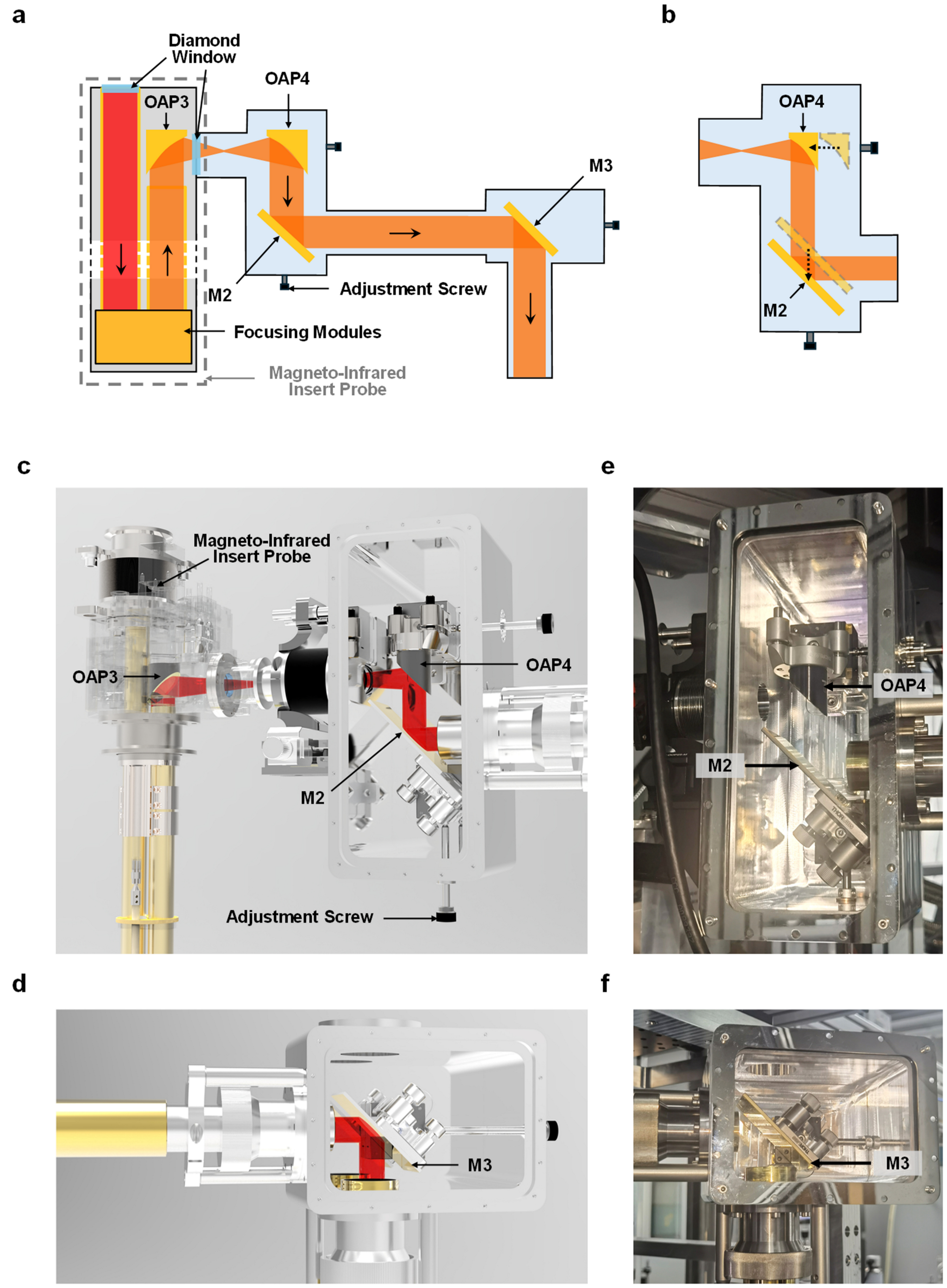

**Fig. 3** Design and photo of the exit collimation chamber. The signal beam emerging from the insert probe is recollimated by OAP4 and redirected by flat mirror M2 into a gold-plated horizontal light tube with polished inner walls. The beam then enters a second vacuum chamber, where mirror M3 directs it downward into another vertical light tube that guides the beam into the detector vacuum chamber. External adjustment screws on OAP4 and M2 enable optimization of beam collimation and direction control. The composite light tube design ensures high reflectivity and vacuum integrity while minimizing polarization degradation.

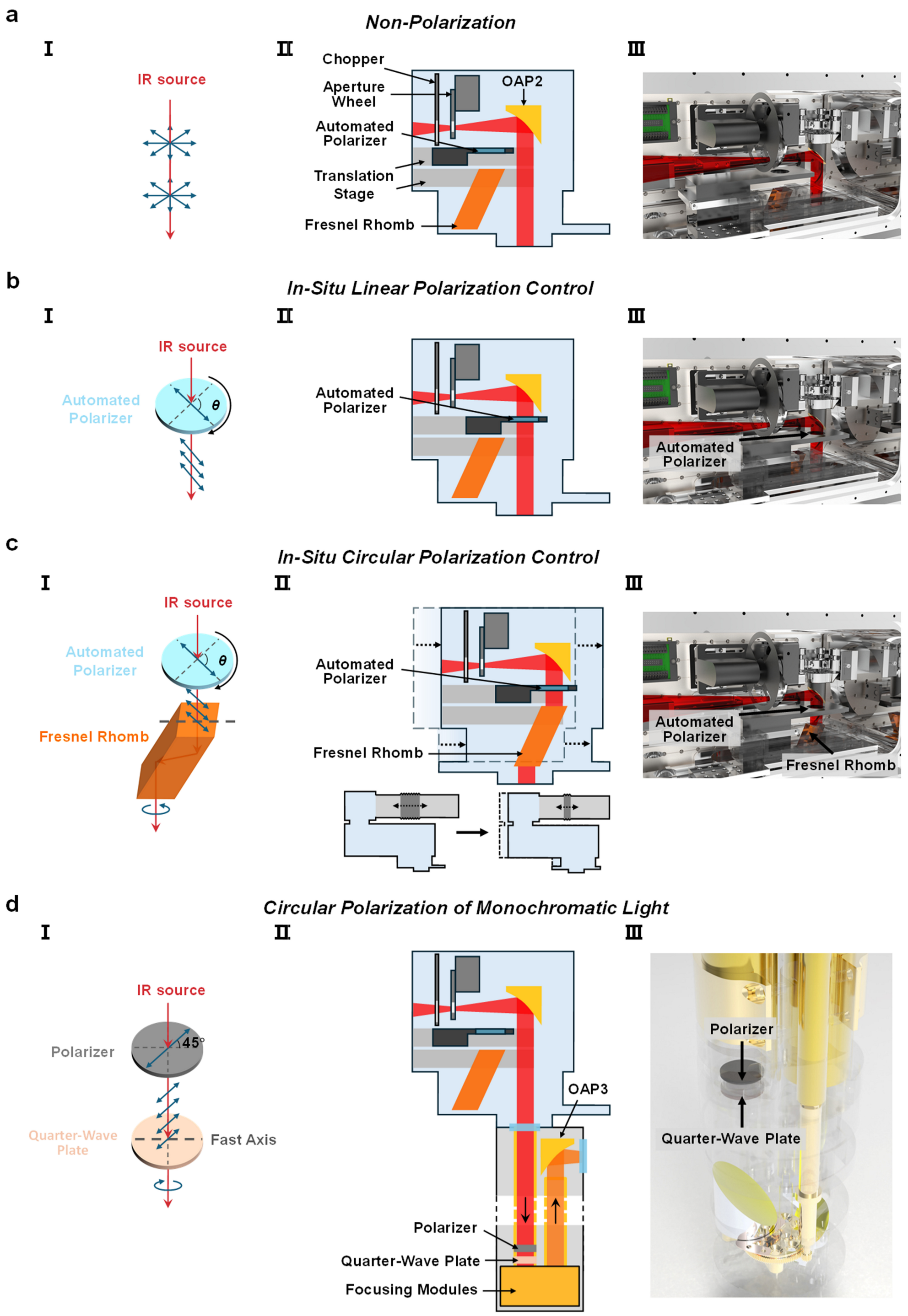

**Fig. 4** The in-situ polarization module for different configurations. Schematic diagrams illustrate the three operational modes enabled by motorized translation stages: non-polarization mode, in-situ linear polarization mode, and in-situ circular polarization mode. A complementary monochromatic configuration is also supported, where polarization optics installed near the sample at low temperature provide maximum polarization fidelity at a specific wavelength.

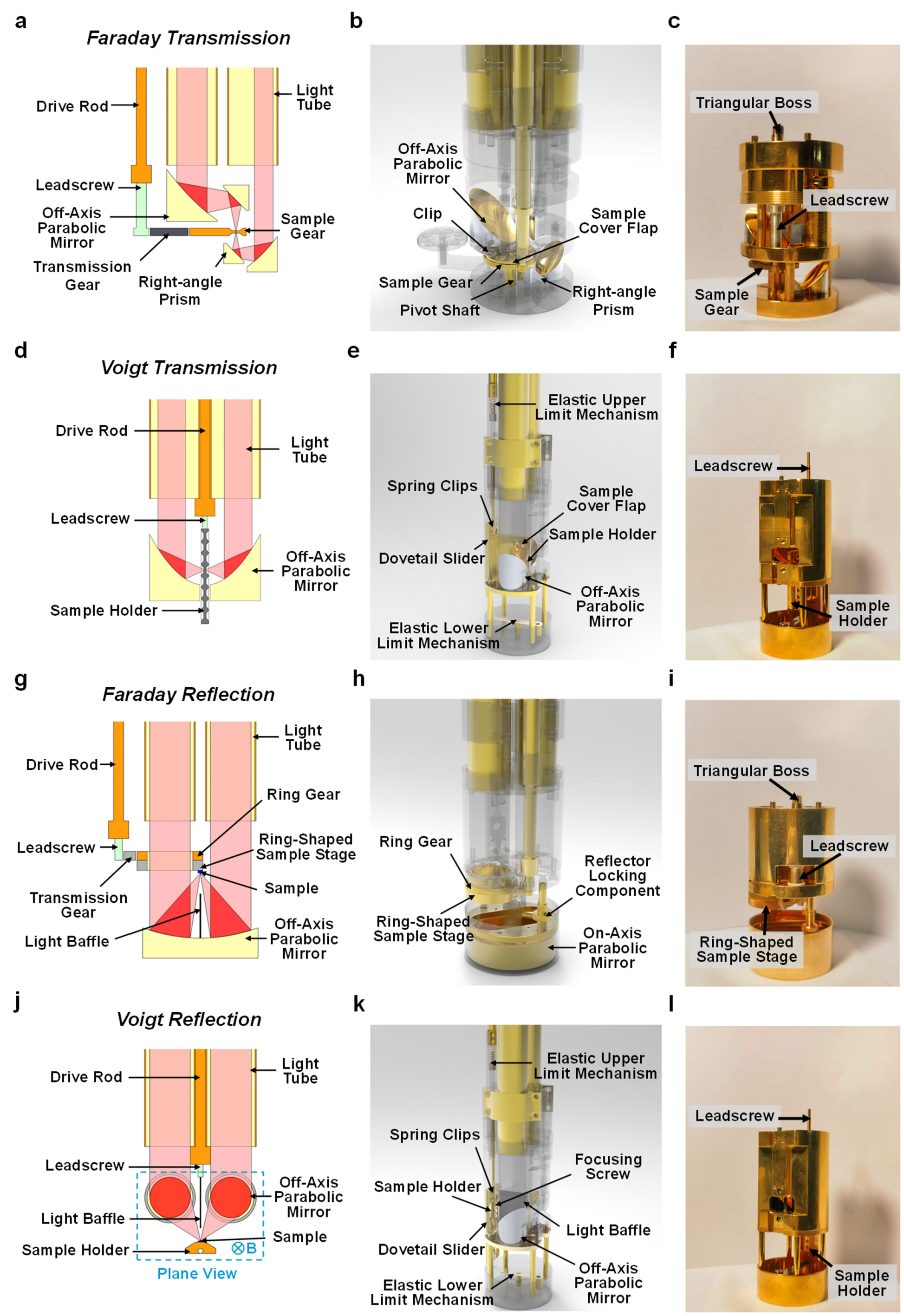

**Fig. 5** Magneto-infrared focusing modules for transmission and reflection geometries. Schematic layouts and key components of the four interchangeable modules: Faraday transmission module, Voigt transmission module, Faraday reflection module, and Voigt reflection module. Each module is designed to fit within the 50 mm bore of the VTI and provides high numerical aperture focusing, low obscuration, high collection efficiency and polarization preservation while accommodating multi-sample holders and mechanical actuation mechanisms.

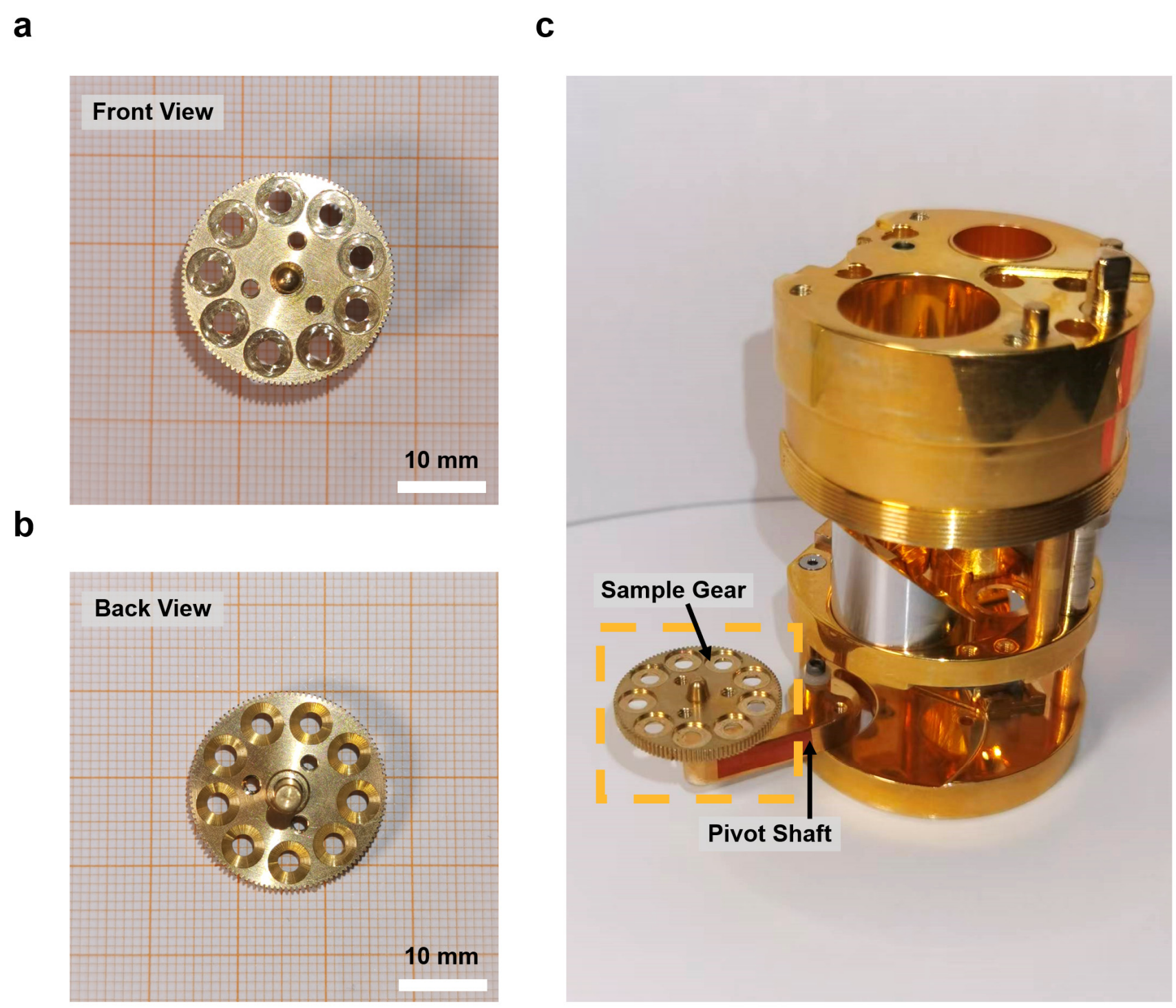

**Fig. 6** Sample gear in the Faraday transmission focusing module. Front and back views of the sample gear, which holds up to nine samples. The sample gear is mounted on a pivot shaft, allowing it to be swung out of the optical path for convenient sample loading. Once the sample gear is returned to its working position, the pivot shaft is mechanically secured by a screw to ensure positional stability.

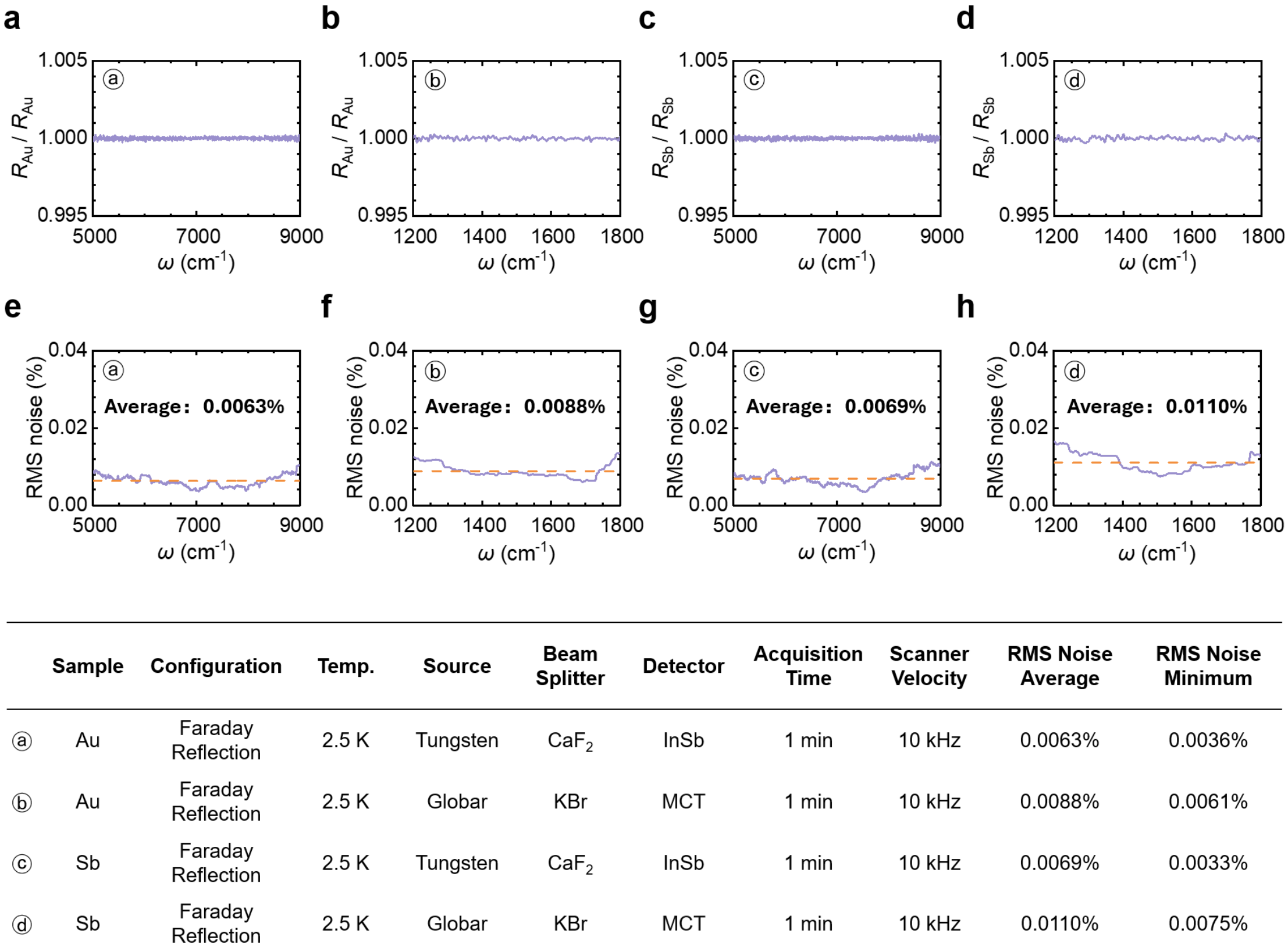

| | Sample | Configuration | Temp. | Source | Beam Splitter | Detector | Acquisition Time | Scanner Velocity | RMS Noise Average | RMS Noise Minimum |
|---|---|---|---|---|---|---|---|---|---|---|
| ⓐ | Au | Faraday Reflection | 2.5 K | Tungsten | $CaF_2$ | InSb | 1 min | 10 kHz | 0.0063% | 0.0036% |
| ⓑ | Au | Faraday Reflection | 2.5 K | Globar | KBr | MCT | 1 min | 10 kHz | 0.0088% | 0.0061% |
| ⓒ | Sb | Faraday Reflection | 2.5 K | Tungsten | $CaF_2$ | InSb | 1 min | 10 kHz | 0.0069% | 0.0033% |
| ⓓ | Sb | Faraday Reflection | 2.5 K | Globar | KBr | MCT | 1 min | 10 kHz | 0.0110% | 0.0075% |

**Fig. 7** RMS noise characterization of the system. (a-d) Noise spectra obtained by taking the ratio of two consecutive measurements on Au film and Sb crystal under zero magnetic field, $R_{Au}/R_{Au}$ and $R_{Sb}/R_{Sb}$, in the configurations listed in the accompanying table. (e-h) Corresponding RMS noise within spectral window of 200 cm$^{-1}$. Orange dashed lines represent the average noise level across each spectral range.

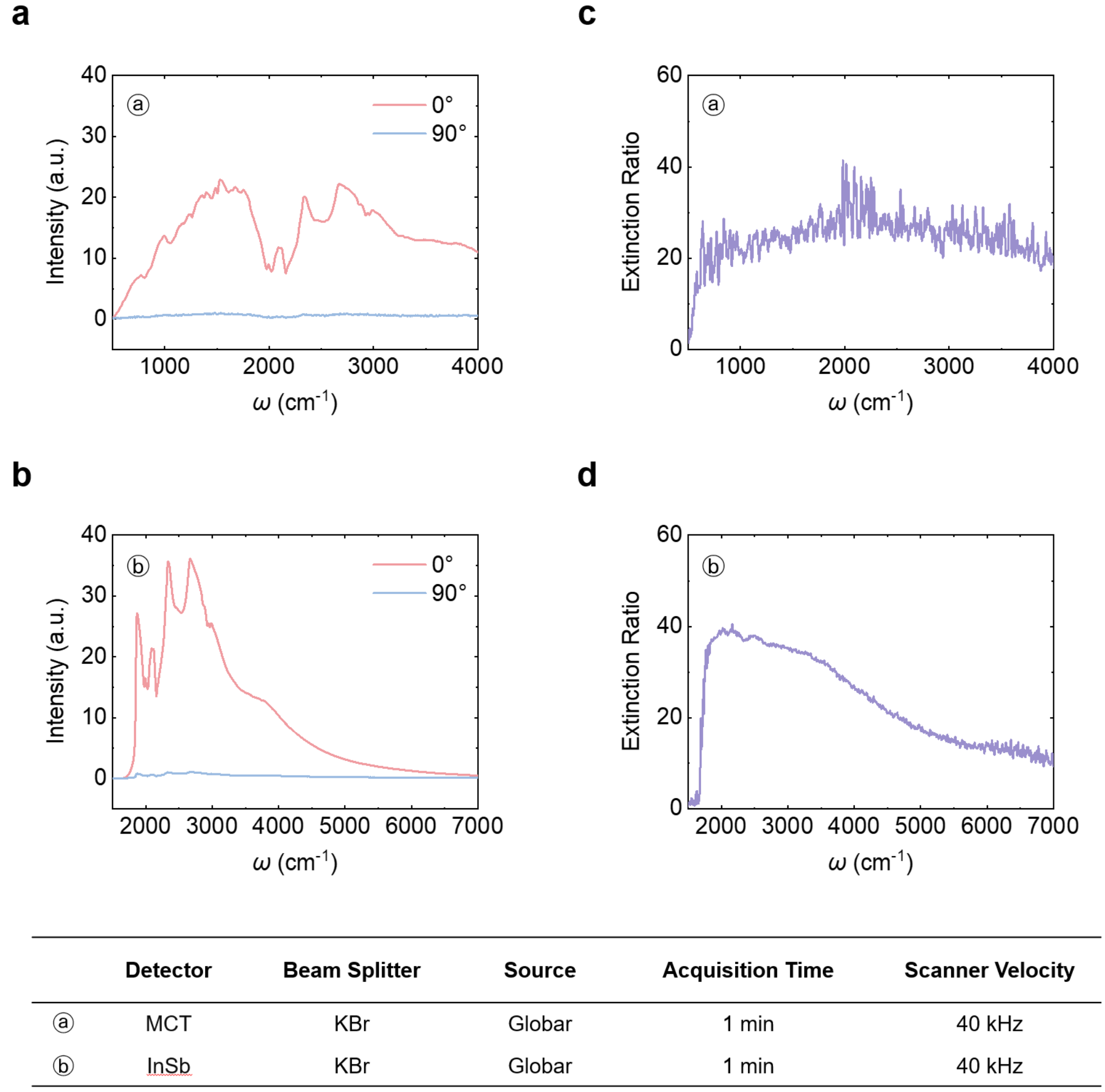

| | Detector | Beam Splitter | Source | Acquisition Time | Scanner Velocity |
|---|---|---|---|---|---|
| ⓐ | MCT | KBr | Globar | 1 min | 40 kHz |
| ⓑ | InSb | KBr | Globar | 1 min | 40 kHz |

**Fig. 8** Extinction ratio measurements in linear polarization mode. (a, b) Intensity spectra recorded with MCT and InSb detectors for parallel (0°) and cross (90°) polarization. (c, d) Extinction ratios.

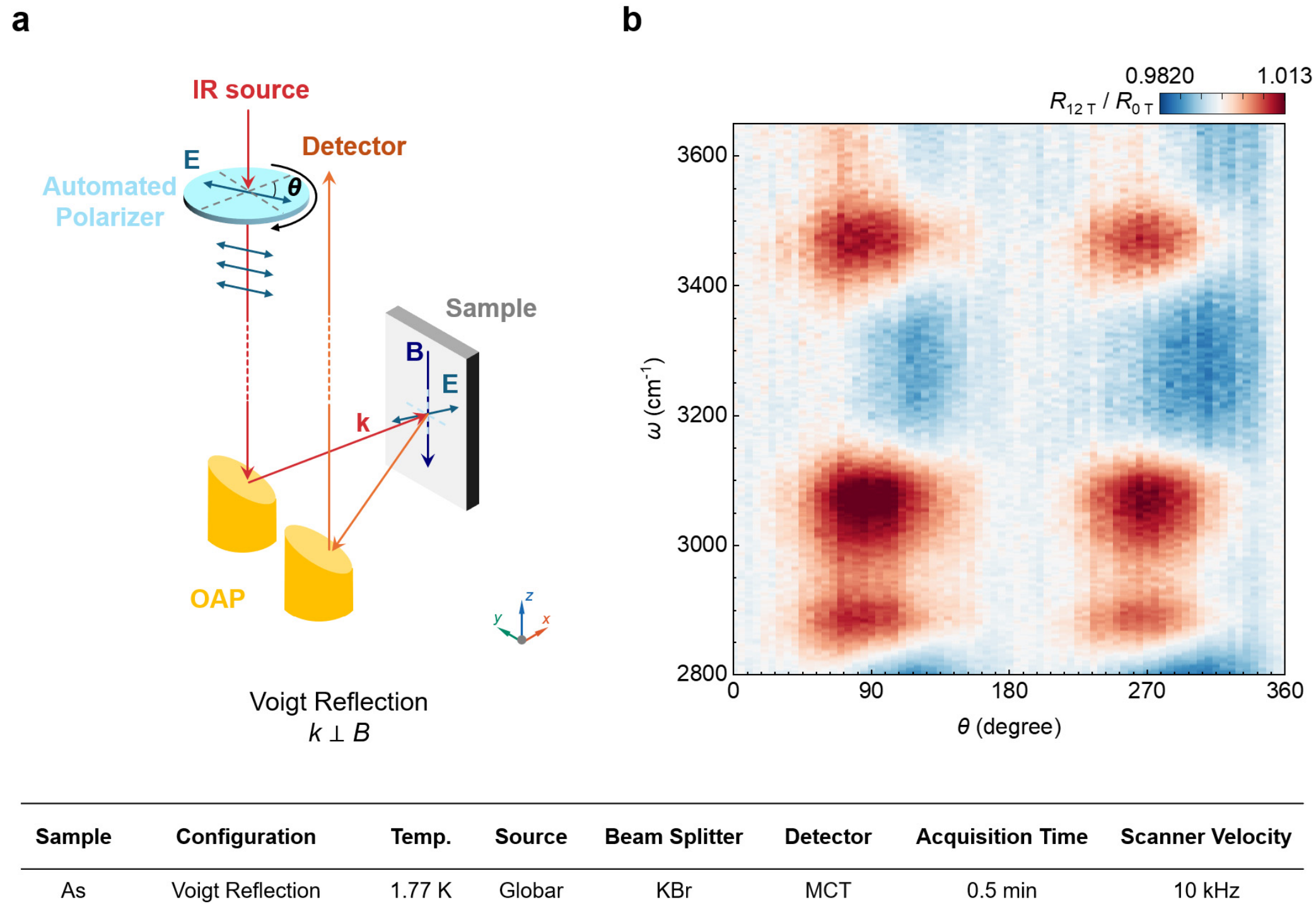

| Sample | Configuration | Temp. | Source | Beam Splitter | Detector | Acquisition Time | Scanner Velocity |
|---|---|---|---|---|---|---|---|
| As | Voigt Reflection | 1.77 K | Globar | KBr | MCT | 0.5 min | 10 kHz |

**Fig. 9** In-situ linear-polarization-resolved spectra of the arsenic crystal in Voigt reflection geometry. Schematic of the Voigt reflection configuration with continuous rotation of the incident linear polarization angle $\theta$ from 0° to 360°, in steps of 5°, and the corresponding relative reflectivity spectra $R_{12\ \mathrm{T}}/R_{0\ \mathrm{T}}(\theta)$ as a function of linear polarization angle $\theta$.

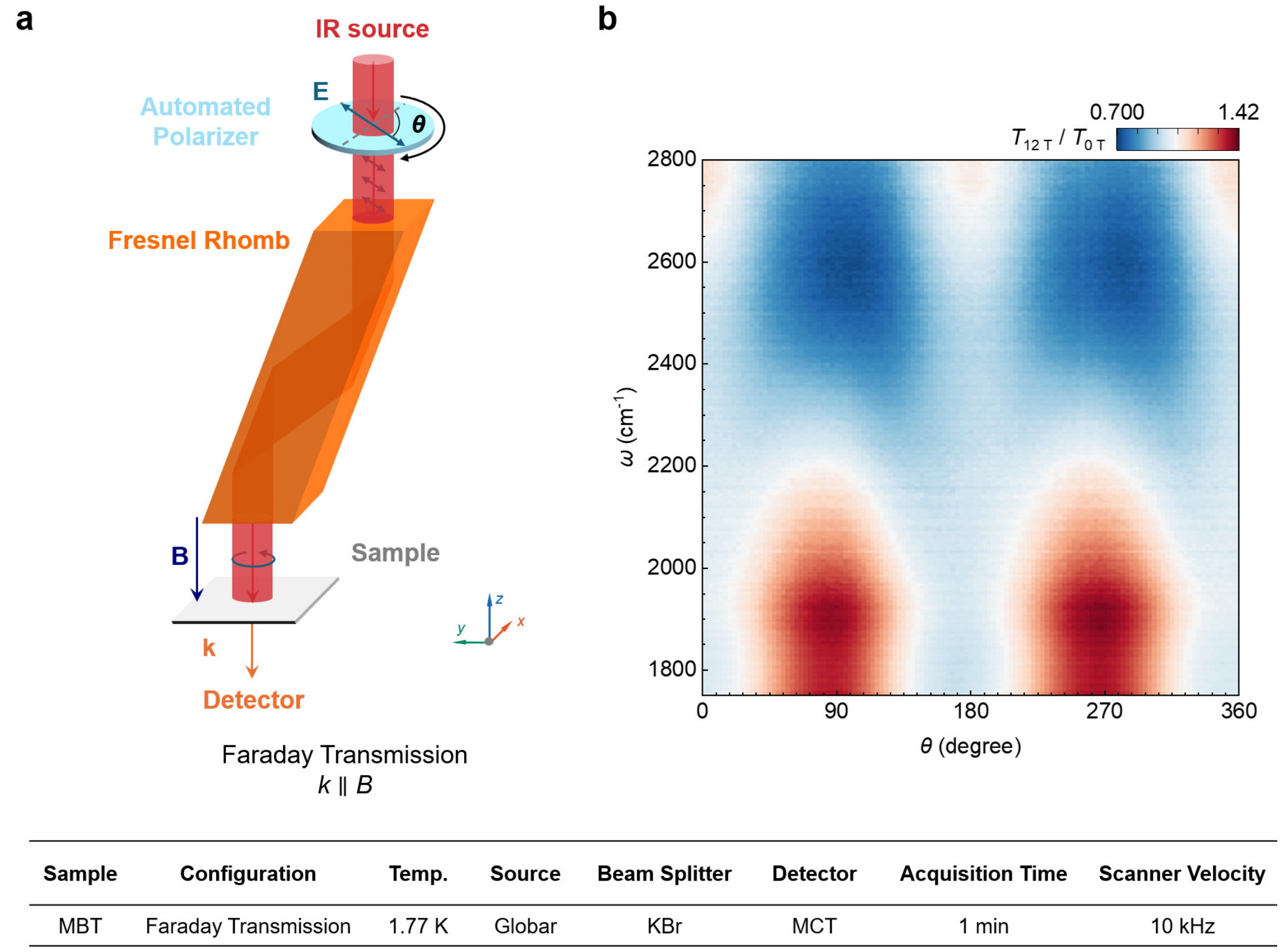

| Sample | Configuration | Temp. | Source | Beam Splitter | Detector | Acquisition Time | Scanner Velocity |
|---|---|---|---|---|---|---|---|
| MBT | Faraday Transmission | 1.77 K | Globar | KBr | MCT | 1 min | 10 kHz |

**Fig. 10** In-situ circular-polarization-resolved spectra of MBT in Faraday transmission geometry. Schematic of the Faraday transmission configuration with continuous polarization modulation, and the measured relative transmission spectra $T_{12\,\mathrm{T}}/T_{0\,\mathrm{T}}(\theta)$.

**Table 1** Reflection statistics for the conventional direct-focusing scheme and the present collimated scheme in this article.

| | $D$ | $\alpha_{max}$ | $N = 0$ | $N = 1$ | $N = 2$ | $N \geq 3$ |
|---|---|---|---|---|---|---|
| Conventional Scheme | — | 7.125° | 0.32% | 2.57% | 5.15% | 91.96% |
| Present Scheme | 3.0 mm | 1.146° | 18.82% | 65.60% | 15.58% | 0.00% |
| | 2.5 mm | 0.955° | 27.47% | 68.02% | 4.51% | 0.00% |
| | 2.0 mm | 0.764° | 41.91% | 58.09% | 0.00% | 0.00% |
| | 1.5 mm | 0.573° | 54.33% | 45.67% | 0.00% | 0.00% |
| | 1.0 mm | 0.382° | 74.62% | 25.38% | 0.00% | 0.00% |